\documentclass[conference]{IEEEtran}
\IEEEoverridecommandlockouts

\usepackage{cite}
\usepackage{amsmath,amssymb,amsfonts}
\usepackage{algorithmic}
\usepackage{graphicx}
\usepackage{textcomp}
\usepackage{xcolor}

\usepackage{url,booktabs}

\def\BibTeX{{\rm B\kern-.05em{\sc i\kern-.025em b}\kern-.08em
    T\kern-.1667em\lower.7ex\hbox{E}\kern-.125emX}}
\begin{document}

\title{MRI2Speech: Speech Synthesis from Articulatory Movements Recorded by Real-time MRI}

\author{\IEEEauthorblockN{Neil Shah\IEEEauthorrefmark{1}\IEEEauthorrefmark{2}, Ayan Kashyap\IEEEauthorrefmark{1}, Shirish Karande\IEEEauthorrefmark{2} and Vineet Gandhi\IEEEauthorrefmark{1}}

\IEEEauthorblockA{\IEEEauthorrefmark{1} CVIT, IIIT Hyderabad, India \mbox{ } \mbox{ } \IEEEauthorrefmark{2} TCS Research Pune, India\\
Email: \{neilkumar.shah, ayan.k\}@research.iiit.ac.in, shirish.karande@tcs.com, vgandhi@iiit.ac.in}}

\maketitle

\IEEEpubid{
    \begin{minipage}{\textwidth}
    \scriptsize
    \raggedright
    Copyright 2025 IEEE. Published in ICASSP 2025 – 2025 IEEE International Conference on Acoustics, Speech and Signal Processing (ICASSP), scheduled for 6-11 April 2025 in Hyderabad, India. Personal use of this material is permitted. However, permission to reprint/republish this material for advertising or promotional purposes or for creating new collective works for resale or redistribution to servers or lists, or to reuse any copyrighted component of this work in other works, must be obtained from the IEEE. Contact: Manager, Copyrights and Permissions / IEEE Service Center / 445 Hoes Lane / P.O. Box 1331 / Piscataway, NJ 08855-1331, USA. Telephone: + Intl. 908-562-3966.
    \end{minipage}
}

\IEEEpubidadjcol

\begin{abstract}
Previous real-time MRI (rtMRI)-based speech synthesis models depend heavily on noisy ground-truth speech. Applying loss directly over ground truth mel-spectrograms entangles speech content with MRI noise, resulting in poor intelligibility. We introduce a novel approach that adapts the multi-modal self-supervised AV-HuBERT model for text prediction from rtMRI and incorporates a new flow-based duration predictor for speaker-specific alignment. The predicted text and durations are then used by a speech decoder to synthesize aligned speech in any novel voice. We conduct thorough experiments on two datasets and demonstrate our method's generalization ability to unseen speakers. We assess our framework's performance by masking parts of the rtMRI video to evaluate the impact of different articulators on text prediction. Our method achieves a $15.18\%$ Word Error Rate (WER) on the USC-TIMIT MRI corpus, marking a huge improvement over the current state-of-the-art. Speech samples are available at \url{https://mri2speech.github.io/MRI2Speech/}
\end{abstract}

\begin{IEEEkeywords}
real-time MRI, silent-speech interfaces.
\end{IEEEkeywords}

\section{Introduction}
\label{sec:intro}
\IEEEpubidadjcol
The vocal tract is crucial in human speech production, shaping the quality and acoustic attributes of generated sounds. This complex anatomical structure, including the glottis, epiglottis, pharyngeal wall, velum, tongue, and lips, undergoes dynamic transformations during speech production. Decoding these transformations into speech—a process known as articulatory speech synthesis—can significantly enhance computer-assisted language learning systems,  provide insights into the precise positioning of articulators concerning the inferred speech, and support individuals with dysarthria \cite{otani23_interspeech}. The challenge lies in accurately estimating vocal tract configurations and correlating them with acoustic representations due to the complexity of the vocal folds, difficulty accessing articulatory movements, and their rapid changes.

Previous research has investigated articulatory synthesis using various sensors and imaging techniques, including  Ultrasound Tongue Imaging (UTI)~\cite{toth2018multi}, Electromagnetic Articulography~\cite{horn1997reliability}, Permanent Magnet Articulography (PMA)~\cite{gonzalez2017direct}, Electrophysiology~\cite{kapur2018alterego}, Electrophysiology~\cite{kapur2018alterego}, Electrolarynx~\cite{espy1998enhancement,kikuchi2004development} and Electropalatography~\cite{kimura2021mobile}. While some of these methods are invasive, they manage to capture articulatory data at high sampling frequencies. However, none of these techniques provide a comprehensive overview of human speech production anatomy in motion. In contrast, rtMRI offers complete coverage of key articulatory structures, such as the hard palate, pharynx, epiglottis, velum, and larynx. Its high spatial resolution allows for in-depth exploration of the physiological basis of speech, singing, emotions, and other aspects of speech production \cite{narayanan2004approach,bresch2008seeing,zhang2010magnetic,ramanarayanan2018analysis,toutios2019advances,belyk2023open}.

\IEEEpubidadjcol
The main challenge in converting rtMRI data to speech is the machine-induced noise present in the ground-truth speech. Despite this noise, most current methods rely on it for training and follow a two-stage approach. First, they extract acoustic features such as MGC-LSP or mel-cepstrum from silent rtMRI videos using CNN-LSTM-based architectures~\cite{csapo20_interspeech,tanji21_interspeech}. Then, these features are used to synthesize speech using HiFi-GAN and WaveGlow decoders~\cite{yu2021reconstructing, otani23_interspeech}. However, relying on ground-truth mel-spectrograms that include speaker-specific details and ambient noise forces models to learn irrelevant information, which reduces intelligibility (WER up to $102.6\%$~\cite{otani23_interspeech}) and hinders generalization. Recent direct single-stage approaches~\cite{wu23k_interspeech} face similar challenges, additionally requiring extensive speaker-specific preprocessing of rtMRI video data. Despite this, error rates remain high (Character Error Rate (CER) at $69.2\%$), and their effectiveness across multiple speakers and datasets remains yet to be seen.

\IEEEpubidadjcol
Alternatively, a learning paradigm can be developed to establish a robust correlation between rtMRI videos and ground-truth text~\cite{pandey2021silent}. This approach offers numerous applications, such as enabling interaction with public displays and kiosks and assisting people with speech disorders, muteness, or blindness in inputting text and interacting with various computer systems. However, relying solely on text as the output modality may overlook the importance of temporal alignment, which is crucial to accurately detect the positioning of specific articulators for language learning.

To address the aforementioned challenges, we propose a novel method called MRI2Speech. Given the inherent noise in the ground-truth speech, achieving high intelligibility through audio modeling is challenging. Therefore, MRI2Speech relies on ground-truth (video, text) pairs to fine-tune a self-supervised audio-visual model, enabling it to infer text from silent rtMRI video input. To ensure accurate temporal alignment specific to each speaker, we introduce a novel duration predictor trained on ground-truth (audio, text) pairs. Using the inferred text and estimated temporal durations, we then train a speech decoder to synthesize aligned speech. We thoroughly evaluate MRI2Speech on the USC-TIMIT MRI~\cite{narayanan2014real} and ArtSpeech Database 1 (ASD1)~\cite{isaieva2021multimodal} datasets. Our method significantly improves rtMRI video-to-speech synthesis, achieving a minimal CER of $9.27\%$-$9.31\%$ and a WER of $15.18\%$-$15.25\%$ on the USC-TIMIT MRI corpus. We also assess MRI2Speech's generalization to unseen speakers across both datasets, with the predicted text yielding a minimal CER of $9.5\%$-$14.66\%$ and a WER of $11.71\%$-$23.11\%$. We assess our model's predictions by examining how, despite masking lip regions in silent rtMRI video, the model focuses on other vocal tract articulators for accurate text prediction, given its pretraining on extensive lip video data.

\begin{figure}[t]
  \centering
  \includegraphics[width=0.63\linewidth]{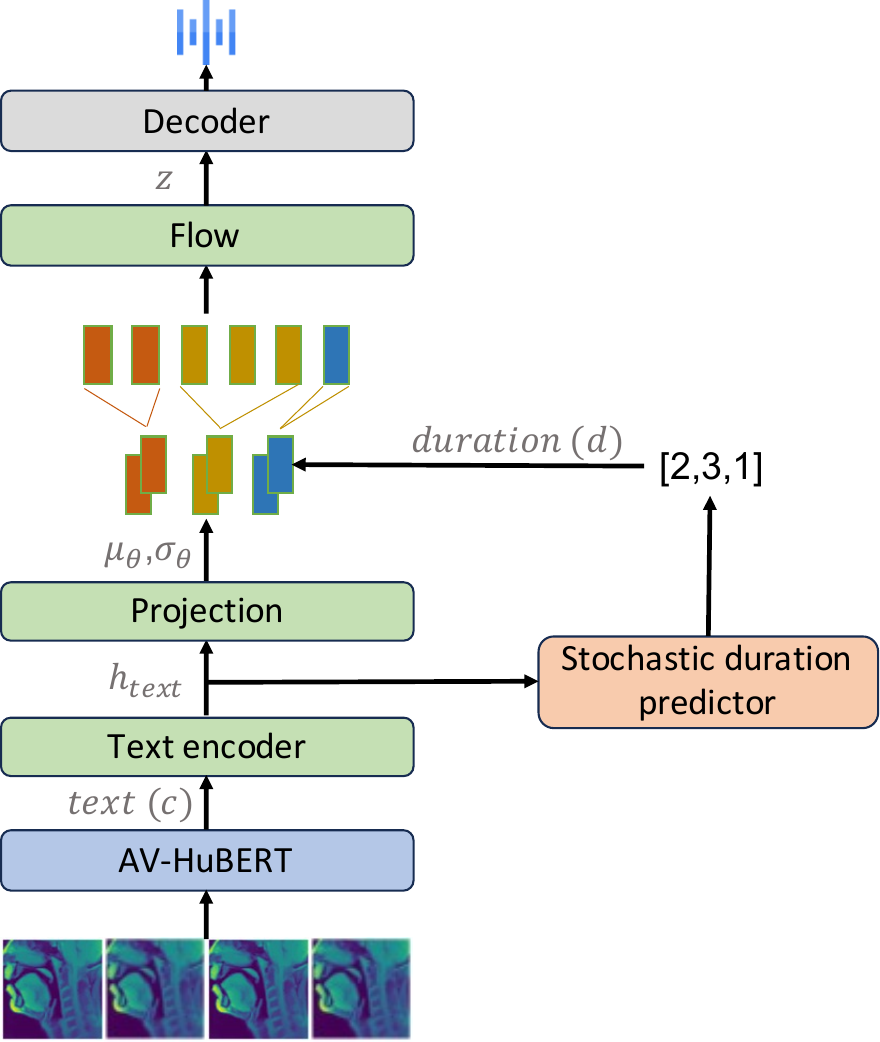}
  \caption{System diagram of the inference process: AV-HuBERT predicts text from silent rtMRI videos. A trained duration predictor expands the phoneme representations, which are mapped to the acoustic space using a normalizing flow, allowing the decoder to synthesize speech.}
  \label{fig:outline}
\end{figure}

\section{Method}
\label{sec:method}

From a system perspective, our main contributions are two-fold: first, we exploit a multi-modal self-supervised model to directly predict text from rtMRI video; second, we estimate speaker-specific temporal alignments by training a duration predictor using available noisy ground-truth audio and text. This allows us to generate aligned speech, either in the speaker's voice or in any desired voice.

\subsection{Fine-tuning AV-HuBERT for text prediction}
\label{sec:estimating content}

Recent developments in self-supervised learning \cite{hsu2021hubert,shi2022learning} have led to high-fidelity, compressed speech representations that emphasize content-rich, disentangled features, excluding speaker and ambient information. This approach simplifies training and supports various downstream applications in producing high-quality transcriptions. AV-HuBERT~\cite{shi2022learning}, a SOTA multi-modal self-supervision model, has demonstrated excellent transferability to lip-reading tasks, achieving strong results with just $30$ hours of paired data from the LRS3~\cite{afouras2018lrs3} dataset. The AV-HuBERT comprises a video and audio encoder, a transformer encoder, and a cluster prediction head. Initially, it conducts feature clustering using audio-based mel-frequency cepstral coefficients to create discrete frame-level targets. In multi-iteration pretraining, features learned from the AV-HuBERT transformer network refine cluster generation in subsequent iterations. 

Given a rtMRI (video, text) pair, we fine-tune the entire AV-HuBERT model to perform visual speech recognition using Connectionist Temporal Classification (CTC)~\cite{graves2006connectionist} loss. We remove the cluster prediction head from the pre-trained AV-HuBERT and replace it with a randomly initialized projection layer that maps the transformer features to labels. The audio encoder is removed to eliminate speech dependence, and its output is replaced with a zero vector. The video features from the video encoder are concatenated to the zero vector. Let the ground-truth transcription have $L$ labels. Consider the input video sequence $x$ of length $T$, we define a AV-HuBERT network with $m$ inputs, $n$ outputs, and weight vector $w$ as a continuous map $\mathcal{N}_w=(\mathbb{R}^m)^T \mapsto (\mathbb{R}^n)^T$. Let $e=\mathcal{N}_w(x)$ be the sequence of visual feature output from the pre-trained AV-HuBERT and $e^t_k$ denote the activation of output unit $k$ at time $t$, which can be interpreted as the probability of observing a specific label $k$ at time $t$. CTC loss maximizes the sum of probabilities for all label sequences leading to the target, as defined in \cite{graves2006connectionist}:
\begin{equation}
    p(k|e^{t})=\sum_{\pi \in \beta^{-1}(k)} p(\pi|e^{T}).
\end{equation}

We use CMUDict~\cite{cmudict} for the lexicon, Adam~\cite{kingma2014adam} as the optimizer with a learning rate that peaks at $0.001$ after $30\%$ of updates, and continue fine-tuning until the validation loss converges ($45K$ steps). For CTC decoding, we employ a pre-trained $4$-gram language model from \cite{shi2022learning} to convert tokens into text.

\subsection{Training VITS for duration estimation}
\label{sec:estimating durations}
We follow the VITS~\cite{kim2021conditional} training paradigm to train a stochastic duration predictor (SDP) for each speaker in both the USC-TIMIT MRI and ASD1 datasets. VITS functions as a conditional variational autoencoder, designed to maximize the log-likelihood $p_{\theta}(x|c)$, where $x$ is the ground-truth audio and $c$ represents the text. The method involves learning specific distributions such as $p_{\theta}(x|z)$, an approximate posterior distribution of generating speech from the latent $z$, and $p_{\theta}(z|c)$, the prior distribution of the latent given text condition $c$. The prior encoder is a transformer-based text encoder, while the posterior encoder consists of WaveNet residual blocks. The HiFi-GAN speech decoder reconstructs the audio from the posterior latents using an adversarial loss. The normalizing flow maps the latent space across both the text and acoustic distributions. A Monotonic Alignment Search (MAS) optimizes the alignment based on log-likelihood scores from the flow decoder. Finally, the SDP estimates the durations for each input phoneme based on these alignments. We train a multi-speaker model using noisy audio-text pairs from both the evaluated corpus. Once trained, the frozen SDP weights from the multi-speaker model are used to predict durations for each input phoneme based on the provided speaker label.

\subsection{Speech Synthesis}
\label{sec:synthesizing speech}
Fig.~\ref{fig:outline} provides a detailed overview of our method during the inference phase. For a given silent rtMRI video, we use our fine-tuned AV-HuBERT model to predict the corresponding text. The text is phonemicized and processed through the prior encoder to generate hidden representations. These representations are then passed through the trained SDP to estimate speaker-specific phoneme durations. The prior latents from the prior encoder are expanded according to these durations and mapped to the acoustic space using the inverse flow. Finally, these mapped representations are fed into the speech decoder to synthesize speech. 

To synthesize speech in a novel target voice, we train a single-speaker VITS model using clean samples from the LJSpeech~\cite{ljspeech17} dataset. The following approach allows for speech generation in a clean voice while preserving the duration alignments of the original noisy source speaker. The source speaker can be any individual from our multi-modal VITS trained on noisy MRI pairs. To generate speech with the source speaker's alignment but in an LJSpeech target voice, we perform inference upon the LJSpeech-trained VITS model and replace its predicted durations with those of the source speaker. The prior latents for the LJSpeech target model are then expanded according to the source speaker's durations. Using inverse flow and the LJSpeech speech decoder, we can synthesize speech in the LJSpeech's speaker voice while preserving the alignment learned from the source speaker. 

\section{Datasets}
{\bf USC-TIMIT MRI dataset:} It includes $3.05$ hours of paired audio, video, and text, containing $460$ sentences spoken by ten American English speakers ($M1$-$M5$ and $F1$-$F5$) while lying supine in an MRI scanner~\cite{wrench2000multichannel}. The audio is sampled at $20 kHz$, and the video resolution is $68 \times 68$ pixels at $23.18$ frames-per-second (fps). Adaptive signal processing algorithms were used to suppress background noise in the audio recordings from the MRI scanner~\cite{bresch2006synchronized}. Given that our approach does not involve audio for learning intermediate text representations, we did not employ additional noise removal techniques as recommended in \cite{otani23_interspeech}. We resized the video frames to $96 \times 96$ pixels and increased the frame rate to $25$ fps to match the AV-HuBERT input requirements. 

\noindent {\bf ASD1 dataset:} It is a subset of a high-quality real-time 2D MRI database introduced in~\cite{isaieva2021multimodal}. It features 77 sentences narrated by ten healthy French native speakers ($P1$-$P10$). The speech data, totaling $1.02$ hours, was recorded simultaneously, denoised, and aligned with text and video. The images were captured at $50$ fps with a $136 \times 136$ pixel resolution and $20$ ms temporal resolution. The larynx is not visible in three speakers ($P8$-$P10$), leading some works to exclude them~\cite{ribeiro2023automatic}. Since these speakers still show valuable articulatory movements, we include them in our evaluation. For each dataset, we reserve $10\%$ of the samples from each speaker for the test set, using the remaining samples for training.

\begin{table}[t]
\caption{Recognition performance of predicted text on USC-TIMIT MRI and ASD1 database.}
\begin{center}
    \begin{tabular}{ccccl}
    \toprule
    Train database & Test database & CER $\downarrow$ & WER $\downarrow$ \\
    \midrule
    USC-TIMIT & USC-TIMIT & 10.95 & 14.38 \\
    ASD1& ASD1 & 11.14 & 22.80 \\
    \midrule
    USC-TIMIT + ASD1 & USC-TIMIT & 27.01 & 40.72 \\
    & ASD1 & 30.67 & 51.11 \\
    \bottomrule
    \end{tabular}
\label{tab:comparison_text_uscasd1}
\end{center}
\end{table}
  
\begin{table}[t]
\caption{Understanding the influence of vocal cord articulators on linguistic learning}
\begin{center}
  \begin{tabular}{ccccl}
    \toprule
    Database & Model's input & CER $\downarrow$ & WER $\downarrow$ \\
    \midrule
    & Lip only & 19.48 & 25.50\\
    USC-TIMIT & Masked Lip & 14.38 & 19.04\\
    & Full rtMRI & \textbf{10.95} & \textbf{14.38}\\
    \bottomrule
  \end{tabular}
\label{tab:lip_nolip}
\end{center}
\end{table}
  
\section{Results and Discussion}

\subsection{Recognition performance on predicted text}
Table~\ref{tab:comparison_text_uscasd1} presents the error rates of predicted text from the AV-HuBERT model, utilizing input rtMRI videos. The predicted text achieves minimal error rates on the USC-TIMIT MRI database (CER: $10.95\%$, WER: $14.38\%$) and on the ASD1 database (CER: $11.14\%$, WER: $22.8\%$). Joint training with speakers from both datasets and testing on individual datasets leads to some drop in performance. However, the CER remains around $30\%$, demonstrating the model's effectiveness in overcoming challenges related to diverse recording conditions, hardware, lighting, language, and head pose.

We mask portions of the MRI videos to analyze how different vocal fold articulators impact text prediction. Our goal is to determine whether the model's effectiveness primarily relies on lip movements or if it also incorporates features from other articulators, given that the AV-HuBERT model, initially trained on $1,326$ hours of lip-reading data, may exhibit biases or a strong dependence on lip regions. We first trained the model using only the lip movements by cropping and inputting just the lip region from the rtMRI video. Next, we masked the lip region and trained the model using only the other visible articulatory movements, such as the vellum, tongue, and glottis. The results, shown in Table~\ref{tab:lip_nolip}, reveal that internal articulators have a more significant impact than the lip regions alone. Inference using only internal articulators (with the lip region masked) results in a $25.33\%$ decrease in WER compared to inference using only the cropped lip region as MRI input.

\begin{table}[t]
\caption{Recognition performance of predicted text on unseen speakers of USC-TIMIT MRI and ASD1 database.}
\begin{center}
  \begin{tabular}{ccccl}
    \toprule
    Method & Database & Unseen speaker & CER $\downarrow$ & WER $\downarrow$ \\
    \midrule
    Ours & USC-TIMIT & $F3$ & 9.5 & 12.67 \\
    & & $F5$ & 8.95 & \textbf{11.71} \\
    & & $M5$ & 12.97 & 18.5 \\
     \midrule
    Ours & ASD1 & $P1$ & 14.66 & 23.11 \\
    & & $P4$ & 7.97 & 17.31 \\
    & & $P10$ & 7.04 & \textbf{15.54} \\
     \midrule
    \cite{pandey2021silent} w/o LM & USC-TIMIT & - & 41.7 & 45.4 \\
    \cite{pandey2021silent} with LM & USC-TIMIT & - & 39.4 & 42.1 \\
    \bottomrule
  \end{tabular}
\label{tab:comparison_text_unseenspeakers}
\end{center}
\end{table}

Table~\ref{tab:comparison_text_unseenspeakers} shows error rates for zero-shot experiments, where a specific speaker is excluded from training. The table presents the results for three different speakers. On average, across all ten folds, our approach achieves (CER: $13.56\%$, WER: $17.06\%$) on the USC-TIMIT MRI dataset and (CER: $11.75\%$, WER: $21.07\%$) on the ASD1 dataset, consistently delivering impressive performance on both the dataset. Compared to the two models from~\cite{pandey2021silent}—one with and one without language models (LM)—which included text predictions for unseen data from two randomly selected speakers, our approach achieves a substantial reduction in error rates. Specifically, our model reports a $65.58\%$ decrease in CER and a $59.47\%$ decrease in WER compared to their best-performing model with LM. This demonstrates the superior effectiveness of our approach in generating more intelligible text for unseen speakers.

\begin{table}[t]
\caption{Performance comparison of synthesized speech from current SOTA method and ours using USC-TIMIT MRI noisy speech data vs. novel voice from LJSpeech speaker.}
\begin{center}
  \begin{tabular}{ccccl}
    \toprule
    Method & Target voice & CER $\downarrow$ & WER $\downarrow$ \\
    \midrule
    Otani et.al.\cite{otani23_interspeech} & - & - & 102.6 \\
    \midrule
    Ours & $M4$ & 39.64 & 65.65 \\
    & $F5$ & 48.04 & 77.15 \\
    & $M5$ & 44.77 & 76.88 \\
    \cmidrule{2-4}
    & $M4$ (LJSpeech voice) & 9.27 & 15.18 \\
    & $F5$ (LJSpeech voice) & 9.31 & 15.25 \\
    & $M5$ (LJSpeech voice) & 10.78 & 17.32 \\
    \bottomrule
  \end{tabular}
\label{tab:comparison_speech_otaniours}
\end{center}
\end{table}

\begin{figure}[t]
  \centering
  \includegraphics[width=0.98\linewidth]{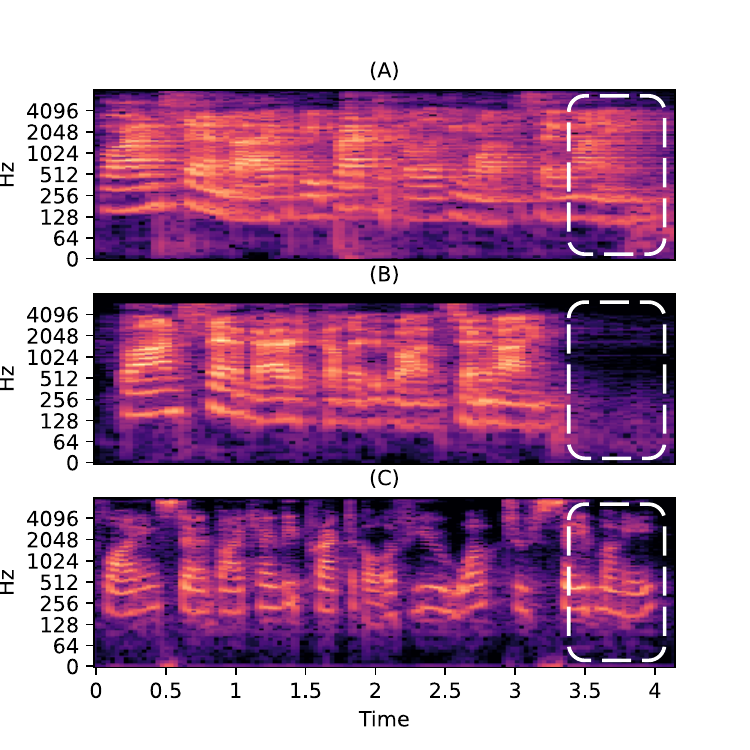}
  \caption{Mel-spectrograms of (A) original speech, and synthesized speech using (B) Otani et al.'s~\cite{otani23_interspeech} approach and (C) our proposed method. The region marked by the white dashed box highlights the inability of current methods to accurately estimate essential formants.}
  \label{fig:spec_otani_ours}
\end{figure}

\subsection{Recognition performance on synthesized speech}
We first synthesize speech using the model trained on noisy audio-text pairs from both datasets to ensure a fair comparison with~\cite{otani23_interspeech}. Table~\ref{tab:comparison_speech_otaniours} presents the error rates for synthesized speech from the USC-TIMIT MRI dataset, specifically for speakers $M4$, $F5$, and $M5$. Our reconstruction with the original $M4$ speaker's voice achieves a $36.04\%$ reduction in WER compared to the current SOTA method~\cite{otani23_interspeech} under similar conditions. However, relying on noisy audio for training may negatively impact performance by causing the model to learn and retain ambient and MRI noise. Additionally, predicting mel-spectrograms limits the model's ability to synthesize speech in different novel voices. To address this, we refine our approach by using original noisy MRI audio solely to obtain speaker-specific alignment using a frozen speaker-specific SDP. We then apply these durations for synthesizing clean speech in the LJSpeech voice (see Section~\ref{sec:synthesizing speech}). This method significantly enhances intelligibility and quality, with WER improving to approximately $15\%$-$17\%$ while maintaining accurate alignment with the video.

Fig.~\ref{fig:spec_otani_ours} compares mel-spectrograms of speech synthesized using~\cite{otani23_interspeech} and our method. Notably, \cite{otani23_interspeech} predicts mel-spectrograms using a speech decoder trained on noisy audio, which affects speech intelligibility as seen by the missing formants in Fig.~\ref{fig:spec_otani_ours} (B). In contrast, our approach (Fig.~\ref{fig:spec_otani_ours}(C)) enhances and preserves fine harmonics, demonstrating robustness even when synthesizing speech from the noisy USC-TIMIT MRI corpus. This advancement in generating high-quality, intelligible speech from rtMRI video, even for unseen speakers, represents significant progress in articulator-to-speech synthesis.

\section{Conclusions}
In this work, we introduce a novel approach to rtMRI video-to-speech synthesis by shifting from predicting acoustic representations to a two-step method. We fine-tune a multi-modal self-supervised AV-HuBERT model for text prediction and train a novel duration predictor for speaker-specific alignments. The predicted text and estimated durations are then fed into a speech decoder to synthesize speech in any novel voice of interest. This shift is necessary because noise in the audio stream can reduce intelligibility and limit the model's ability to synthesize speech in novel voices. As a novel strategy, we demonstrate self-supervision's effectiveness in understanding the ability of articulators internal to the human body for speech production. On the USC-TIMIT MRI corpus, the synthesized speech achieved a WER of $15.18\%$, offering a $575.89\%$ improvement over the existing SOTA method. Our method enables effective generalization to unseen speakers across different rtMRI databases, overcoming challenges like hardware variations, head-pose changes, and lighting differences. Future work will focus on embedding emotive features from articulators into the synthesized speech.

\bibliographystyle{IEEEtran}
\bibliography{refs}
\end{document}